\documentclass[pre]{revtex4-1}

\usepackage{mathrsfs,amsmath}
\usepackage{graphicx}
\usepackage{color}

\newcommand{\be}{\begin{eqnarray}}
\newcommand{\ee}{\end{eqnarray}}

\begin{document}%

\thispagestyle{plain}

\setcounter{page}{1}

%
%
%
%




\title{How log-normal is your country?\\An analysis of the statistical distribution of the exported volumes of products}

\author{Mario  Alberto Annunziata*}

\affiliation{*Corresponding author: marioalberto.annunziata@isc.cnr.it\\CNR- Istituto dei Sistemi Complessi \\
Area della Ricerca Roma Tor Vergata,\\
via del Fosso del Cavaliere 100, 00133 Roma -Italy}

\author{Alberto Petri}
\affiliation{CNR - Istituto dei Sistemi Complessi \\Dipartimento di Fisica, Universit\`a Sapienza,\\
P.le A. Moro 2, 00185 Roma - Italy}

\author{Giorgio Pontuale}
\affiliation{CNR - Istituto dei Sistemi Complessi \\Dipartimento di Fisica, Universit\`a Sapienza,\\
P.le A. Moro 2, 00185 Roma - Italy}

\author{Andrea Zaccaria}
\affiliation{CNR - Istituto dei Sistemi Complessi \\Dipartimento di Fisica, Universit\`a Sapienza,\\
P.le A. Moro 2, 00185 Roma - Italy}

\begin{abstract}
We have considered  the statistical  distributions of the volumes of the different producst  exported by 148 countries. 
 We have found that the form of these distributions  is not unique but
 heavily depends on the level of development of the nation, as expressed by macroeconomic indicators like GDP,
GDP per capita, total export and a recently introduced measure for countries'
economic complexity called fitness.
We have identified three major classes: \textit{a)} an
incomplete log-normal shape, truncated on the left side, for the less
developed countries,
\textit{b)} a complete log-normal,with a  wider
range of volumes, for nations characterized by intermediate economy, and  \textit{c)} a strongly asymmetric 
 shape for countries with a high degree of development. 
The ranking curves of the exported volumes from each country seldom cross each other, showing  a clear hierarchy of export volumes.
Finally, the log-normality hypothesis has been checked for the distributions of all the 148 countries through different tests, Kolmogorov-Smirnov and Cram\'er-Von Mises, confirming that
it cannot be rejected  only for the countries of intermediate economy.
\end{abstract}

%
%

\maketitle

\section{Introduction}
\label{sec-introduction}

The volume of exported goods is a useful, although partial, proxy for assessing the economical status of nations. 
Statistics of exported volumes from different countries have been considered  from several authors under many aspects \citep{easterly09}.
Especially recently, fluxes between exporters and importers have been investigated also from the point of view of complex networks  \citep{fagiolo10,squartini11}.

Here we have considered the gross volume exported in each different sector  by 148 countries 
and found that there is no a unique form characterizing their statistical distribution. Indeed, the shape of the distribution evolves more or less continuously according to the macro indicators characterizing  the development of  nations. As such indicators  we have considered the GDP, the per capita GDP, the total export,
and the fitness, recently introduced in  \citep{newmetrics}. 
Continuous evolution in the distribution shape of nations is  especially well shown  by the ranking  curve of the exported volumes 
of each nation, that we discuss in Sec. 2. Each curve seldom crosses the others and a clear hierarchy of countries is visible.
Concerning the densities, discussed in Sec. 3, less developed countries display distribution shapes close to a log-normal, but truncated on the left side. Distributions widen their range and become complete log-normal for intermediate nations,  whereas for those with high degree of development they show an asymmetric, skewed shape, 
shared by many other phenomena \cite{bee11}.

Finally,  we have checked the log-normality of the distribution of all the 148 countries against different tests, including 
Kolmogorov-Smirnov and Cram\'er-Von Mises, finding not completely agreeing results, as reported in Sec. 4. In any case log-normality must be very likely rejected for
the nations having the smallest and the largest economies. We shortly discuss some possible origins of these differences in Sec. V.

\section{The national ranking of the exported goods}
\label{sec-ranking}
The UN Comtrade database, as processed in \citep{abbracci} contains, among other information, the exported volumes of about  150 countries aggregated for different productivity sectors.  These sectors are labeled by numerical codes whose number of digits expresses the level of aggregation, from a few tenths, classified by two digits, to about five thousands of classes, classified by six digits. We adopted the four-digit classification, including about twelve hundred classes, since in our opinion it represents a good compromise between product identification and excessive detail. Data are available for many
different years. As an instance we shall show here the results for the year 2010, but we have checked that similar results are obtained also for the years between 1995 and 2009.

We have started our analysis by ranking in descending order the exported volumes of each sector for each country in the database. Less developed countries have in general less items than the others, and the ranking curves will be shorter. It is important to remark that, in general, the volume ranking  will list  products in different order for each  country, since there is no strict correspondence between the rank of a volume and the kind of product, although we shall see that some regularities on the large scale can be detected.

\begin{figure}
\includegraphics[width=120mm]{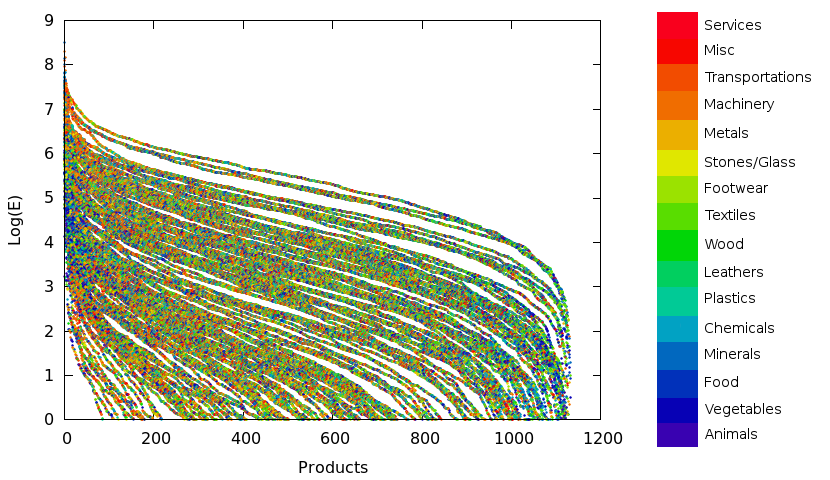}
\caption{The ranking curves of export volumes for the different countries. Each curve starting  above another rarely crosses the curves below. The colors indicate where each product category locates itself in the ranking. Larger exporters clearly differentiate  in the tails, medium in the head ($E$ are in U.S. k$\$$, $\log$ is in base $10$).
\label{fig-1}
}
\end{figure}

The curves resulting from  the ranking are reported in Fig.~\ref{fig-1}, in thousands of U.S. dollars. It can be seen that they are very well separated, and that those belonging bigger exporters lay above those of smaller exporters in the whole range, so 
the first important property that emerges is that  curves seldom cross each other: if the  top ranked volumes of a given country are larger than the top ranked of another, in almost all cases this relation will hold for all the ranked volumes. This unveils a hierarchy, where nations exporting less products have curves staying below those of nations exporting more products. In other words there is a direct correlation between the exported volume of each product and their total number. 

Another important point concerns the appearance of the curves: Not only they lay at different heights in the graph, but also have different shapes. The weak-exporting countries have curves decaying almost linearly in the lin-log plot of Fig.~\ref{fig-1}, implying an exponential decay, whereas  intermediate countries follow a slower decrease, which ends with a  sudden drop in the  countries with the largest volumes, showing that they export low volumes very seldom, as displayed  by the sparsity of point in the tails. 

Although, as remarked above,  the volume ranking does not produce the same order of products in the different curves, some regularities can be caught on a wide scale. The different colors on the curves of Fig.~\ref{fig-1}    
represent different classes of products according to the two-digit Comtrade classification,  reported  in the legend together with the color assigned to each one. It emerges that classes of low  average  technological content appear more frequently in the tail of the curves of big exporters, but in the head for the others, where instead the formers display mostly high-tech products. Of course, these  qualitative considerations require more refined and quantitative analyses in order to draw stricter conclusions, and can be subject of future work .

\section{The frequency histograms of exported volumes: How are they log-normal?}
\label{sec-histograms}
To our knowledge this is the first statistical investigation of the export volumes of a nation aggregated for product and importer, so we have no information about which kind of distributions they might be tight to.
However,  similar  studies have been performed in the past for related  quantities, specifically for the export flows  \citep{fagiolo10}, finding in some cases frequency distributions close to a log-normal. Frequently, log-normals with right-handed power-law tails (also called Pareto-log-normal distributions) have also been found.

Log-normal distributions are observed whenever statistics result from the product of independent,  random variables provided of variance. It is not obvious to which extent the export abilities of a nation can be generated by such combinations. However, there have been cases where observations have   been successfully interpreted in terms of such processes. A historical example comes from the study on the scientific productivity of researchers that the Bell Labs commissioned  to the Nobel laureate  Shockley, the inventor of transistor  \citep{Shockley1957}. 

In his study Shockley discovered that, not only at Bell's, labs have generally a log-normal distribution of scientific production
rates, and explained this occurrence in terms of the set of different capabilities that each researcher has to possess and to put together (multiply) in order to produce a scientific paper or patent. So we can assume, in analogy, that the volume of  each exported good results from joining (multiplying) a certain set of capabilities of the country,  that are possessed in a given amount and that change from good to good. This combinatorial approach resembles the microscopic model illustrated in \citep{zaccaria2014taxonomy}.

Thus, after building a histogram for each country, representing the empirical frequencies of exported  volumes, we have followed the suggestions coming from the theoretical and empirical arguments exposed above, and fitted the export volume  distributions with log-normal distributions. 

As for the ranking, we have performed a thorough study of the export volumes from 148 countries adopting the 4-digit Comtrade indexing of goods.
In this way the  number of products exported from each country ranges from about 100 to 1131. In the present work we report  these analyses for the year 2010, but similar results are generally obtained for the other years we have analyzed, from 1995 to 2009.
For each country we built an histogram representing the empirical probability of exporting a given volume, irrespective of the corresponding products.
Actually our statistics considers the logarithms of the exported volumes and of the resulting frequencies, and it is therefore expected to be parabolic if our assumption of log-normality is true. 

\begin{figure}
			\includegraphics[width=80mm]{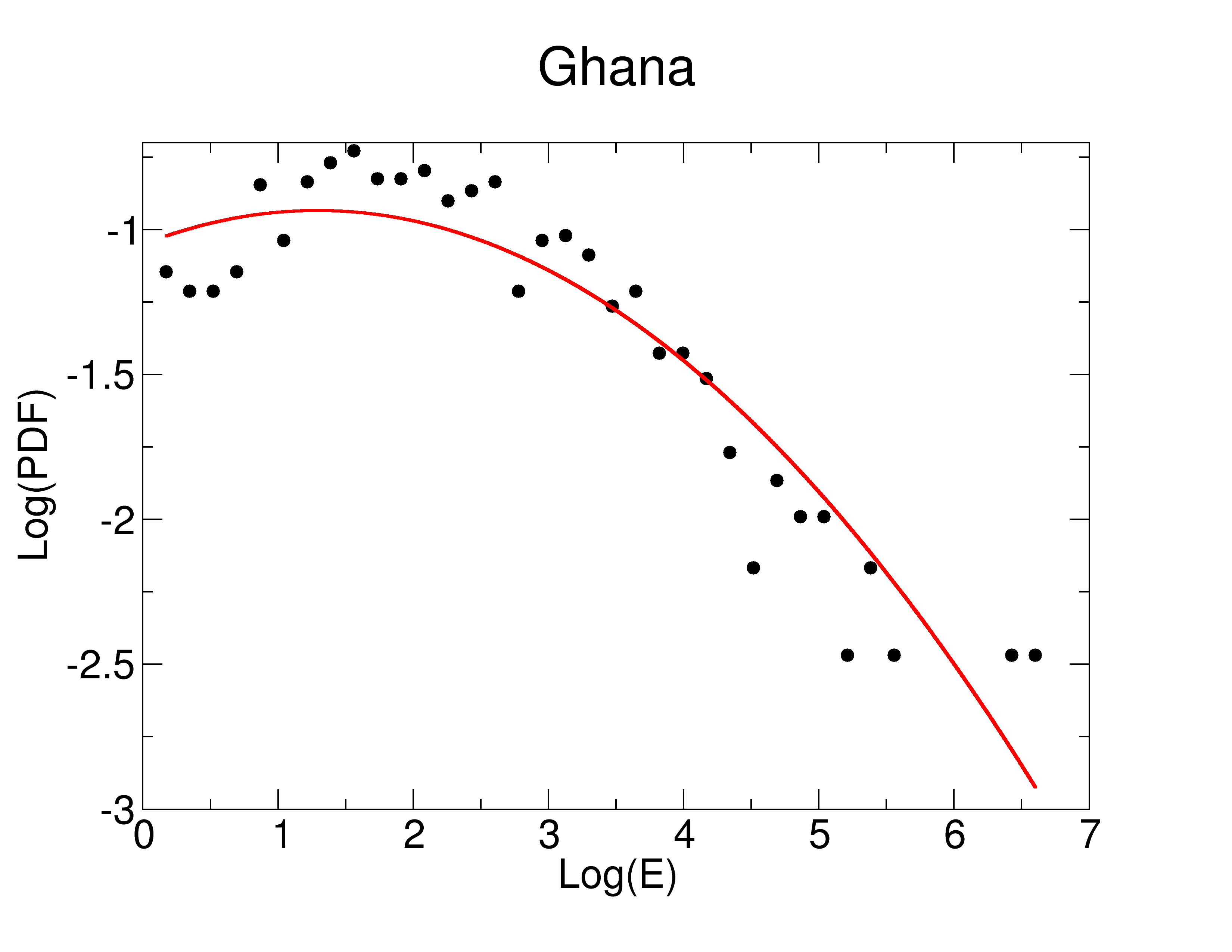}
					\includegraphics[width=80mm]{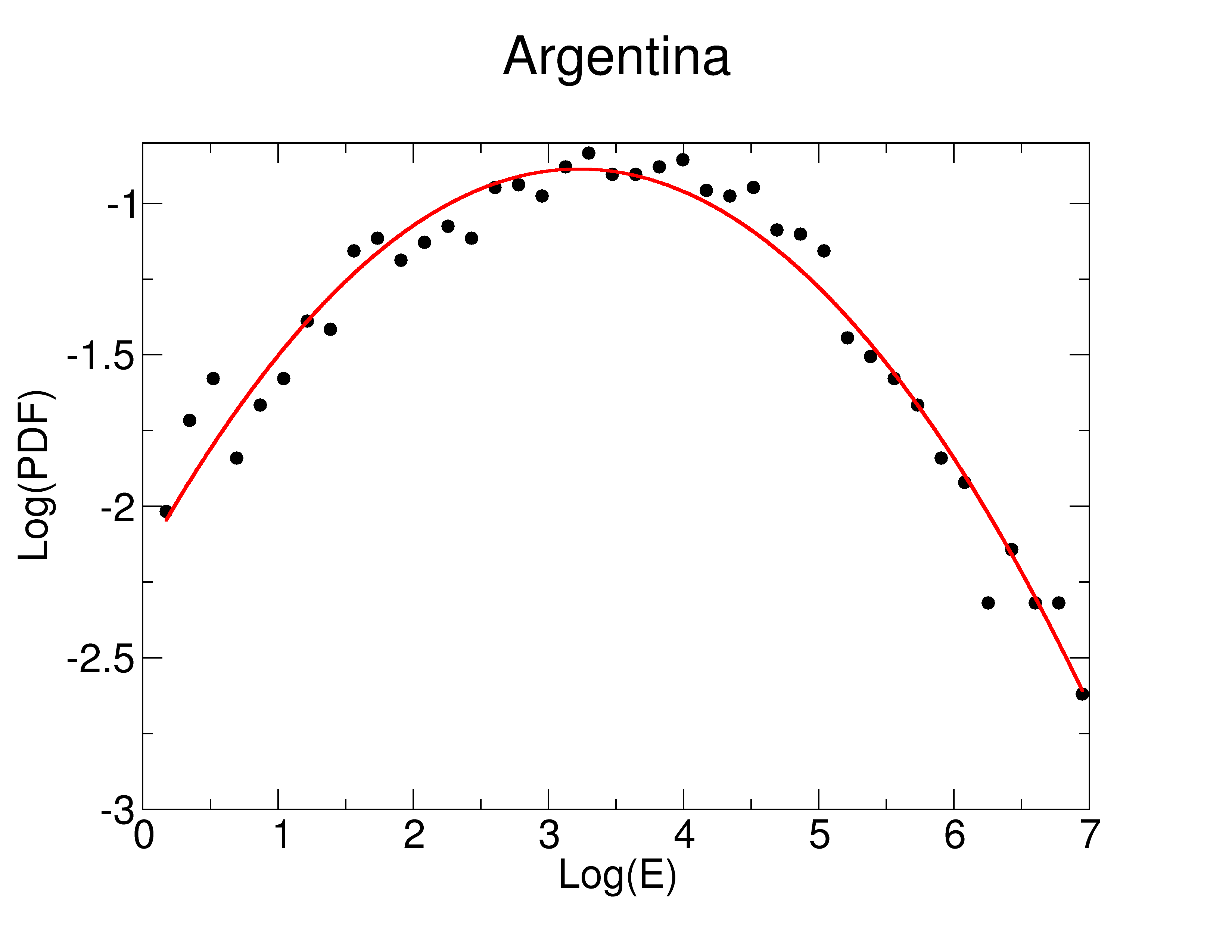}
				\begin{center}
							\includegraphics[width=80mm]{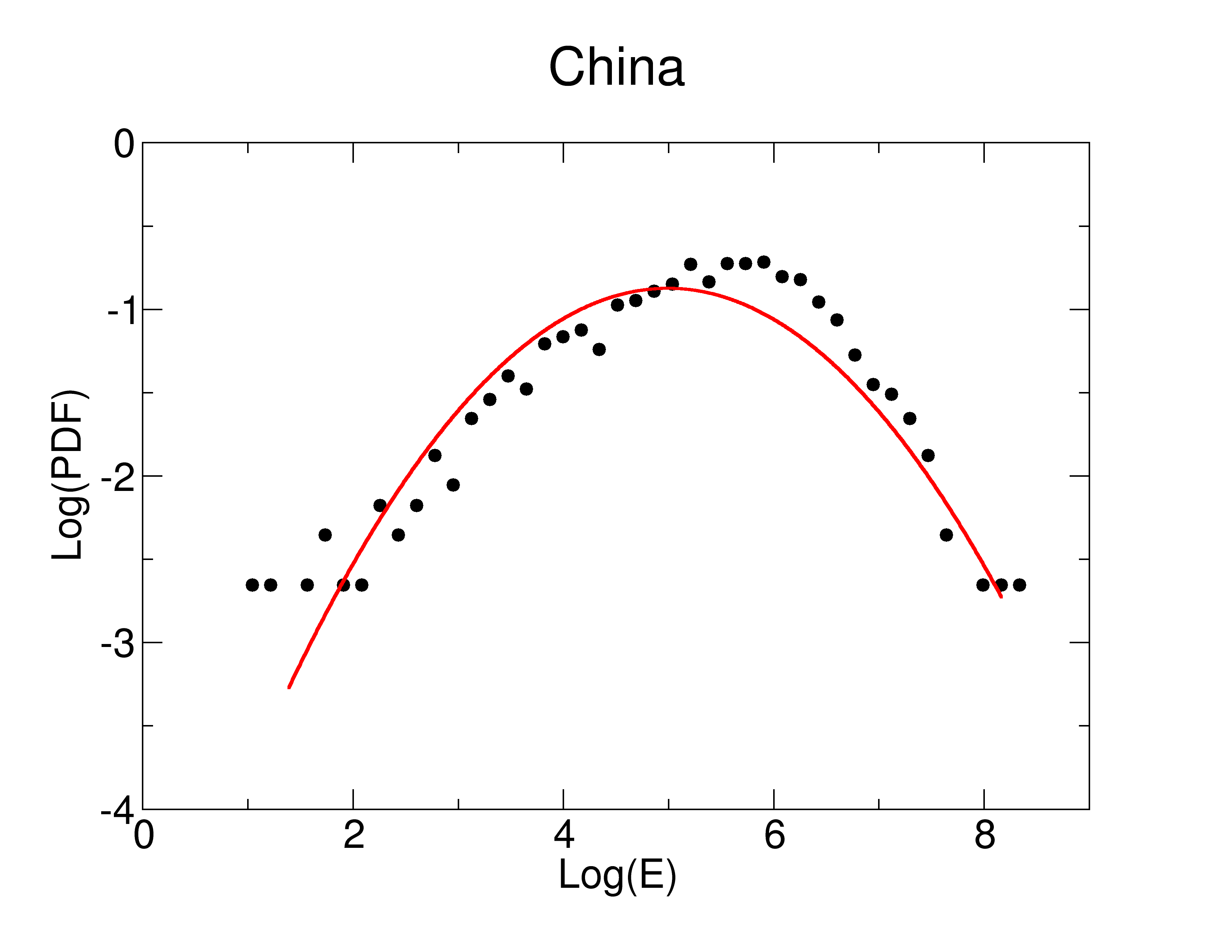}		
				\end{center}
	\caption{Frequency distribution of exported volumes (thousands of dollars)  for three countries representative of the different typologies:  Ghana (up left), Argentina (right) and China (down).  Red solid lines represent log-normal fits (logarithms are in base $10$).}
	\label{fig-3}
\end{figure}

A glance at the shapes of the resulting distributions suggests that they can be roughly grouped into three main categories, that however have no sharp separation. Rather, passing through the statistics of the  different countries, shapes are seen to evolve gradually.
In Fig.~\ref{fig-3} three instances of the different emerging shapes are shown. To the first group belong the countries whose curves lay at  bottom in the graph of Fig. 1, like in this case  Ghana. Their export distribution appears quite close to a log-normal, but  with a cut at low values.
The mode of the distribution is low ($\log E < 3$, where $E$ is the volume in thousands of dollars), and a complete log-normal distribution cannot show itself.
The second group is made by the middle-settled  countries. The shape of their export distribution appears very well fitted by the log-normal, with  larger parameters than  the low-lying nations.
The mode of the  distribution is higher ($ 3 < \log E < 
7$) and a log-normal distribution is fully displayed with standard deviation around $10^3 < \sigma < 10^4$, like for Argentina in the figure.
The world's largest exporters, like China, belong to the last category, lying on top in Fig.\ref{fig-1}, with the largest modes and standard deviations Their distribution is log-normal for the sector involving not too large volumes, but the right hand side of the curve displays a different character, similar to the above mentioned Pareto-log-normal. When trying to fit them with a log-normal,  a "bump"  becomes evident  at the right side of the fitting parabola vertex. However, one can easily check that repeating the fit after dropping the points on the right of the modal value, a log-normal with 
different parameters can be found that  well suits  the empirical data, on the left wing, and the bump disappears. So one can conclude that what actually happens is that the distribution changes shape on the right side.

\section{Empirical evidences on country's features in the light of  macro indicators}
The results of the previous section clearly show that the economic level of a country affects not only the amount of its  export but also its volume distribution. This poses the problem of identifying which factors are responsible for the change not only in the amount, rather in the character of the exported volumes.
This is a very hard challenge, and  in the present work we limit ourselves to check if any  relation between the distribution shape and 
macroeconomic indicators exists. For the latter we have made three choices. 
Three are very simple and widely employed indexes: the Gross Domestic Product, the same per capita (GDPpc)  and the total exported volume. The fourth is a more sophisticated quantity recently introduced with the name of \textit{fitness}  \citep{newmetrics}.
The definition of this  estimator of the competitiveness of a nation, that takes into account the features of the export in a non-linear and self-consistent way, is briefly recalled in the Appendix. It can be useful to note that both GDPpc and fitness are intensive (i.e. per capita) quantities, whereas the total export and GDP are extensive.

In order to highlight the possible relationship between the  indicators and the volume export distributions we have colored each curve of picture of Fig.~\ref{fig-1} on the base of the values that a given indicator assumes for the respective country.  

\begin{figure}
	\begin{center}
		\includegraphics[width=90mm]{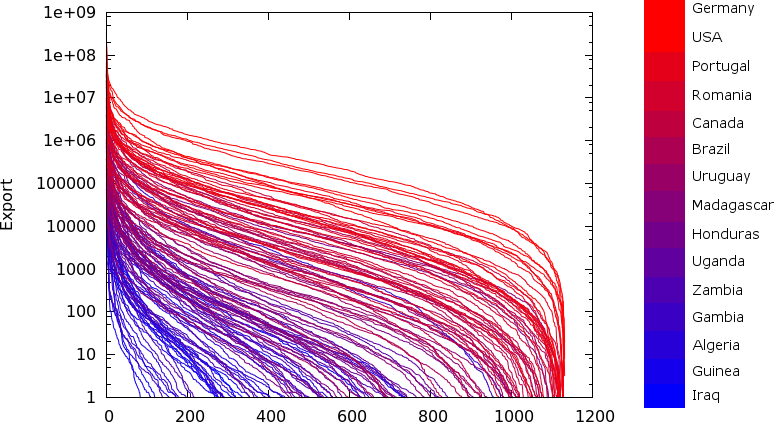}
			\includegraphics[width=90mm]{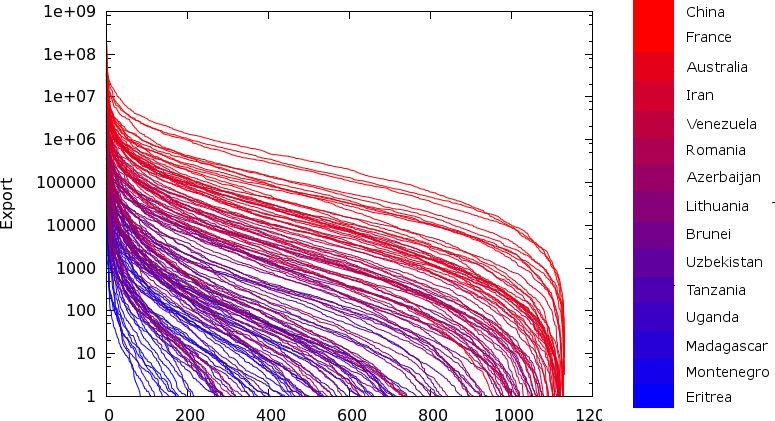}
							\includegraphics[width=90mm]{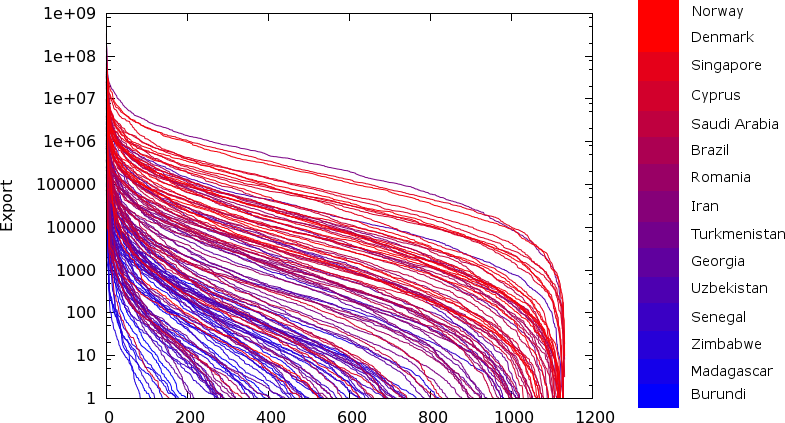}
							\includegraphics[width=90mm]{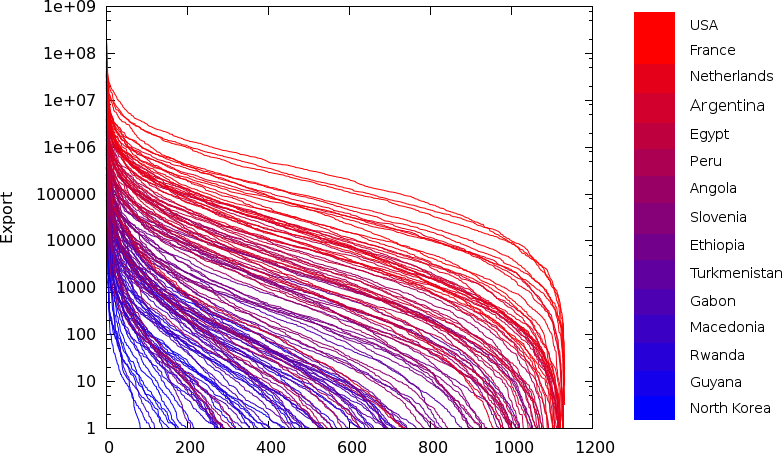}
	\end{center}
	\caption{Ranking curves of exported volumes for each country, colored in each  pictures according to the value of a different indicator. From top to bottom: fitness, total export,  
	GDPpc, and GDP. Blue corresponds to the lowest values, red to the highest.}
	\label{fig-4}
\end{figure}

\begin{figure}
	\begin{center}
		\includegraphics[width=120mm]{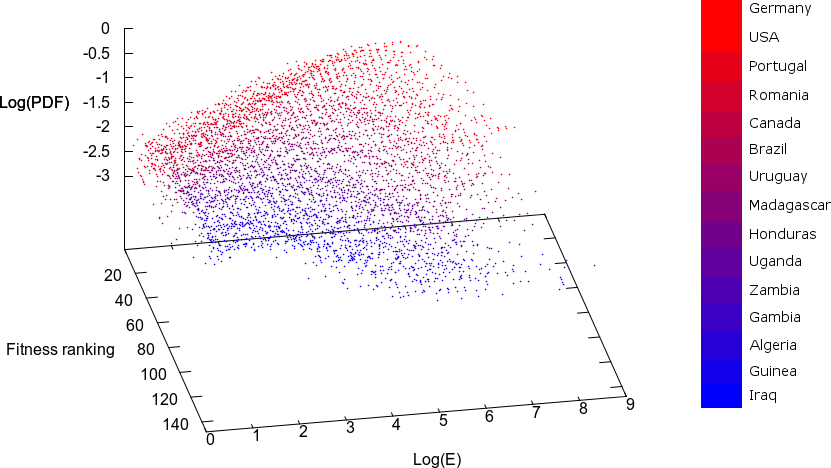}
	\end{center}
	\caption{A three-dimensional plot of the export distributions for the different countries, ordered by the fitness (blue: low fitness, red: high fitness). By increasing the fitness, shape goes from a truncated-log-normal to the fully-log-normal, ending with  Pareto-log-normal (logarithms are in base $10$).}
	\label{fig-5}
\end{figure}

In Fig.~\ref{fig-4}, \ref{fig-4}a),  \ref{fig-4}b) and \ref{fig-4}c) we
report the same ranking curves of Fig.~\ref{fig-1}, this time giving each curve a color, from blue (low values of the macro indicator) to red (high values). Actually, in  order  to get a more linear distribution  of colors, each of them does not refer to the actual value of the indicator  rather to its ranking among the 148 values \citep{plosnm,zaccaria2015case}.
All the indicators appear to be significantly related to the health status of the economy of a country,  as far as this can be associated to how high is the relative position of its export curve in the graph. This happens despite two of them, total export and GDP, are extensive (proportional to the country size), and two are  intensive (fitness and GDPpc). From these plots it is clear that countries with higher indicators export more products and in higher volumes, although the GDPpc, Fig.~\ref{fig-4}b, seems not to be as good as the others.   It is seen that in this case some curves on top may  have low indicators and vice versa.  On the other hand, the pronounced monotonicity of the color gradient for the other indicators, from low values (blue) to high values (red),  suggests  that these  should be more correlated to observed  hierarchy in the countries distributions. If one sorts all the density curves discussed in Sec. \ref{sec-histograms} according, for instance, to the fitness  of the corresponding nation,  one goes in  ascending order from the more truncated-log-normal distributions to the more skewed Pareto-log-normal,
passing through the fully log-normal ones. This is well shown in the $3$-d plot displayed in Fig.~\ref{fig-5}  where all the countries'  export distributions are shown, colored according to the fitness.
A quantitative analysis could get rid of  which indicator is really more correlated with the position of the corresponding ranking curves  with respect to the others, and will be subject of a forthcoming work.

\section{Statistical tests for the  distributions}

 The frequency histograms of the exported volumes discussed in the previous sections offer a good glimpse on the differences characterizing countries of different economies. However, binning data in histograms always suffers from some degree of arbitrariness with respect to the number of adopted bins. 
In order to assess the reliability of the log-normal assumption on the base of statistical tests,  we have thus considered the empirical cumulative distributions of the exported volumes. They are strictly related to the ranking curves, discussed in Sec. 2, from which they can be obtained in a straightforward way by switching the axes and ranking the $N$ volumes of the given country in ascending order from $1/N$ to $1$.
 Thus the shape of the curves remains the same, although the scales change.
 
\begin{figure}
	\includegraphics[width=120mm]{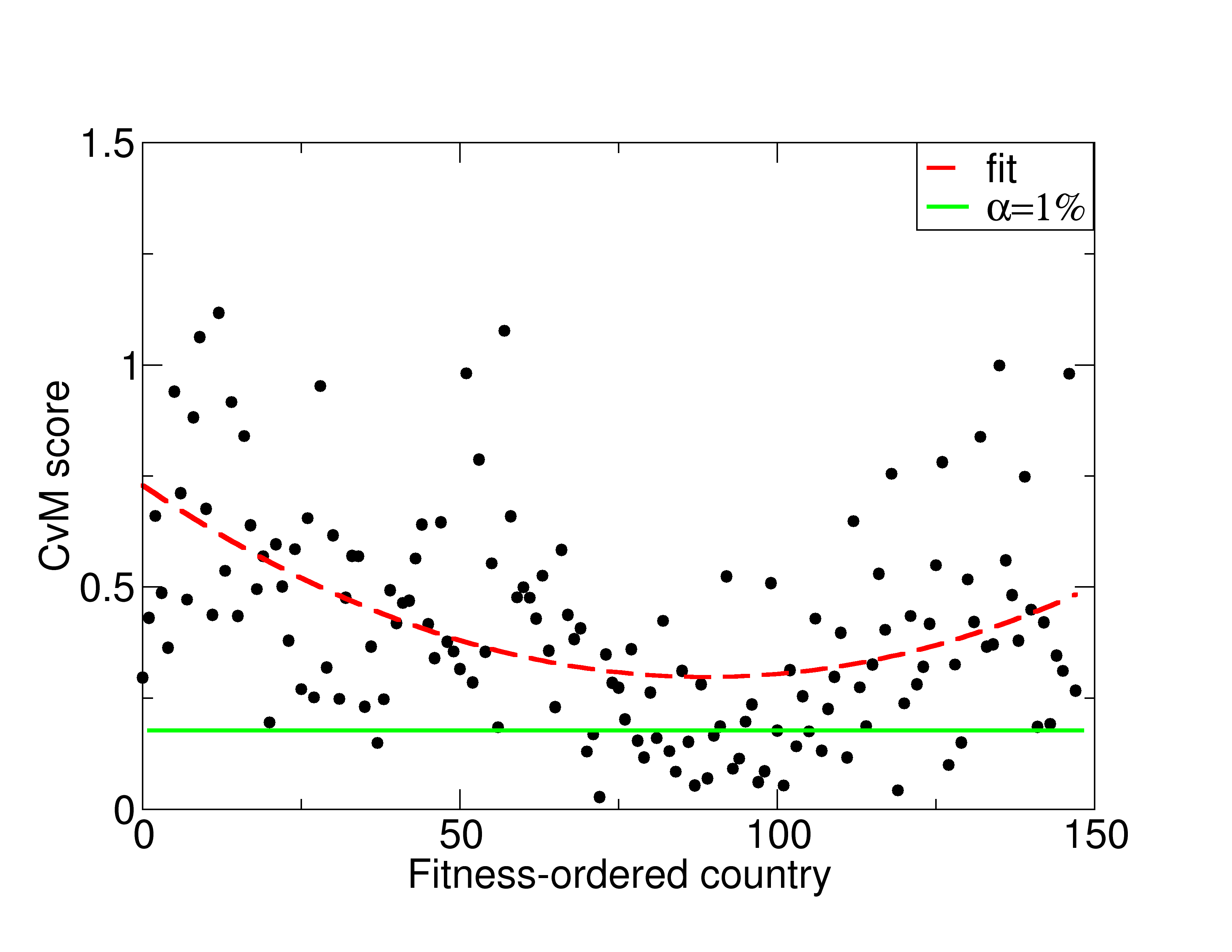}
		\includegraphics[width=120mm]{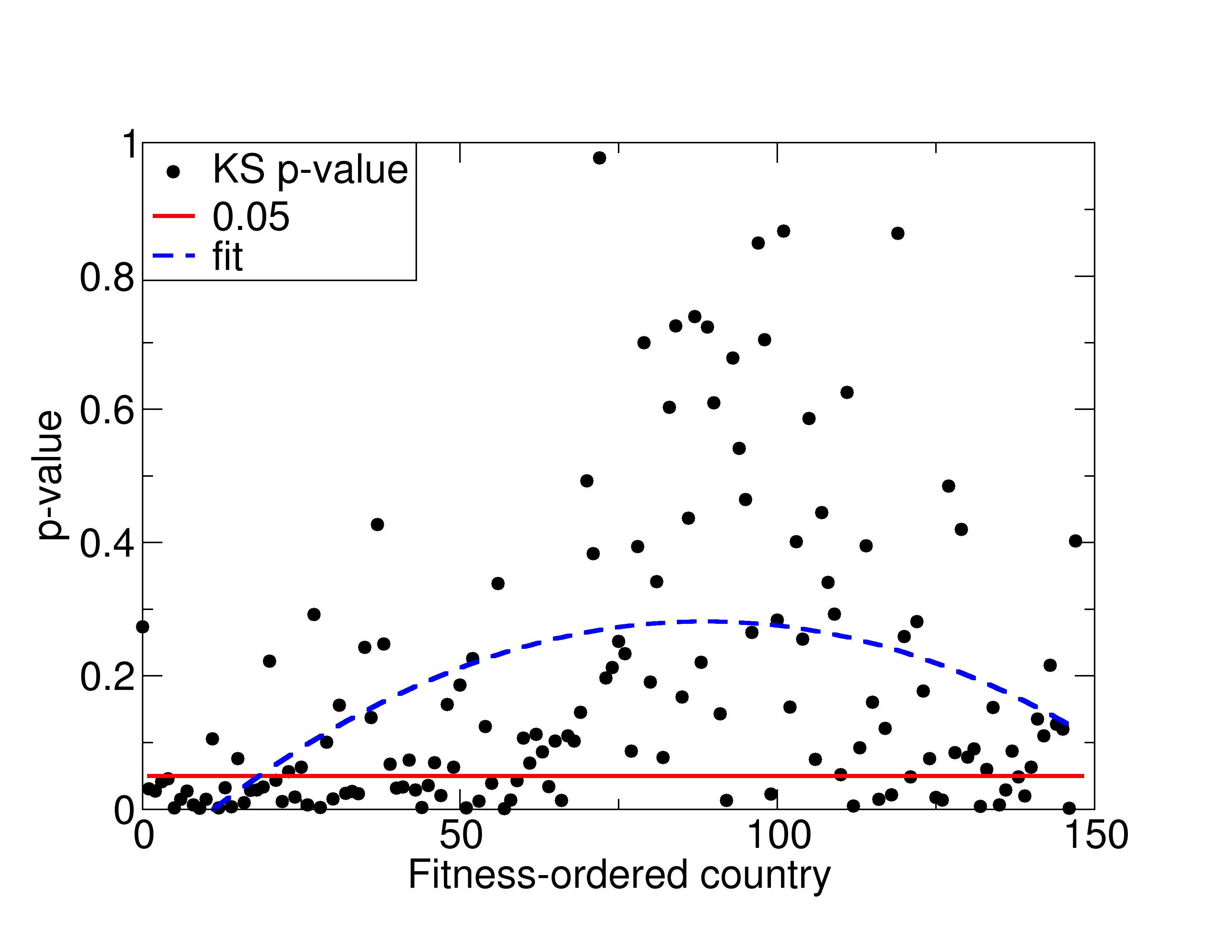}
	\caption{Top: Results of Cr\'amer-von Mises statistics applied to the cumulative distributions of the export of countries, under the hypothesis of log-normality. Green line: rejection threshold corresponding to $\alpha = 1\%$, red dashed line: a parabolic fit of  the  statistics. Bottom: results of the Kolmogorov-Smirnov statistics. Setting acceptance at 5$\%$ log-normal hypothesis must be rejected for the points below the red line. The blue line is a parabolic fit of the obtained probability values}
	\label{fig-6}
\end{figure}

In order to check the goodness of log-normal fits, we performed Cram\'er-von Mises (CvM)   and Kolmogorov-Smirnov (KS) \citep{Darling1957} tests for the cumulative distributions of the exported volumes under the hypothesis of log-normality. For the CvM test  the values of the resulting statistics are such that even with a
  significance as low as $\alpha = 1\%$, log-normality cannot be rejected only for 26 countries out of 148 . This  result is reported graphically in Fig.~\ref{fig-6}a), where  the countries are ordered following the ranking induced by the fitness. It is seen that log-normality can be accepted in the  full-log-normal region (medium fitness), while for distributions belonging to truncated-log-normal (low fitness) and Pareto-log-normal regions (high fitness) the log-normal hypothesis has to be  rejected. The trend of CvM-score is worth notice, because it definitely shows ``how log-normal a country is'': it decreases in the truncated-log-normal region, has a minimum in the full-log-normal region and then increases in the Pareto-log-normal region.
The KS test produces a similar qualitative behavior, shown in Fig.~\ref{fig-6}b). However, the level needed to reject a significant number of countries is sensibly higher, $5\%$ and despite this, the  number  remains sensibly larger than with the CvM test.  The two tests give thus different acceptances for the log-normal hypothesis, but there is in any case a set of countries of intermediate ranking whose volume statistics  can reasonably be assumed log-normal.

\section{Summary, conclusions and perspectives}
\label{sec-conclusion}

The statistics  of the export volumes of countries, aggregated in product categories, has shown   distinctive features that depend 
 strongly on the values of such macroeconomic indicators as national GDP, GDPpc, total export volume, or fitness. Specifically, log-normal curves seem 
 to fit well the distributions for middle and low-developed countries, but with a lacking tail at low volumes for the latter.  Countries with large export display a skewed curve, with a  tail shorter at high volumes
than at low ones. Their shape resembles other statistic sometimes denominated Pareto log-normal.

The log-normal shape characterizing middle economy countries points to some 
multiplicative process  behind the production, and exportation,  of goods.
As sketched above, this could be related to the capabilities needed
to accomplish the goal. Truncation  of small export volumes in less developed countries suggests the existence of a  minimal threshold  to generate effective export.

Several mechanisms have been proposed to explain skewed Pareto log-normal shapes, as well as different approaches for their precise mathematical characterization  \citep{reed04,bee11,melitz2003impact,helpman2003export}.
We just wish to point out that, assuming a genuine multiplicative process at the origin of the observed log-normalities, there  can be  different causes making them to deviate from the normal convergence and for the on-set of asymmetries  at large volumes. The simplest, at least from the mathematical point of view, is the presence of correlations among the variables contributing to the multiplicative process. This can give rise to shapes different from the (log) normal with  rich and interesting properties \citep{janosi1999,petri2008,bramwell02}. On the other hand, the steepest decay of the  tails at high volume could signal  the existence of upper limits to export, due to the buyers demand and resources in the world trade. Hence a full log-normal distribution cannot develop and, on the right of the maximum, volumes  fall  rapidly.

We are confident that further investigation  can relate more quantitatively the economic indicators to the distribution shape and the relative country positioning. This could allow to make assessments about the different economies,  still more if carried on together with studies on the time evolution of the curve shapes and parameters.

\section*{Acknowledgments}
We are grateful to A. Tacchella for supplying  processed Comtrade  data, and to M. Cristelli and E. Pugliese for stimulating  discussions.  A.P. and A.Z.  also thank the European project FET-Open GROWTHCOM (grant num. 611272) and the Italian PNR project CRISIS-Lab.

\section*{Appendix: A primer on the Fitness and Complexity algorithm}
In a recent paper, Tacchella et al. \citep{newmetrics} proposed a novel approach to measure some intangible properties of countries and of the products they export. Their base is constituted by the ideas of Hidalgo and Hausmann \citep{HH} the total exported volume or te{HH}, who introduced a linear, coupled algorithm to measure such properties starting from the export basket of countries. Tacchella et al. prosed an improved, non linear algorithm, whose new features are motivated by empirical \citep{primonewmetrics} and economical \citep{plosnm} evidences. We refer to these references for an exhaustive presentation of such approach, which is beyond the scope of the present contribution. The input data is given by the binary export matrix $M$, whose elements $M_{cp}$ are equal to $1$ if country $c$ has a Revealed Comparative Advantage \citep{balassa1965trade} in exporting the product $p$. The algorithm calculates, by means of two coupled iterative equations, two variables which are proxies for countries' competitiveness, the \textit{Fitness} $F$, and products' complexity, $Q$, as a function of $M$ and the other variable. At each 
step both variables are normalized in such a way that their sum is equal to the total number of countries and products in the database, which correspond to the rows and the columns of the matrix $M$, respectively. In formulas, the algorithm to calculate the fitness $F_c$ of the country $c$ and the complexity $Q_p$ of the product $p$ can be written in the following form:
\begin{equation}
 \tilde{F}_c^{(n)}=\sum_p M_{cp} Q_{p}^{(n-1)}
\end{equation}
\begin{equation}
 \tilde{Q}_p^{(n)}=\frac{1}{\sum_c M_{cp} \frac{1}{F_{c}^{(n-1)}}}
\end{equation}
\begin{equation}
\label{jsdlskjfs}
 F_c^{(n)}=\frac{ \tilde{F}_c^{(n)}}{ <\tilde{F}_c^{(n)}>_c}
\end{equation}
\begin{equation}
\label{ljlkjljlkj}
 Q_p^{(n)}=\frac{ \tilde{Q}_p^{(n)}}{ <\tilde{Q}_p^{(n)}>_p}
\end{equation}
where the normalization of the intermediate tilded variables is made as a second step by dividing each $\tilde{F}_c^{(n)}$ and $\tilde{Q}_p^{(n)}$ by the respective averages,
\[
<\tilde{F}_c^{(n)}>_c \equiv \frac{1}{C} \sum_c \tilde{F}_c^{(n)} \qquad <\tilde{Q}_p^{(n)}>_c \equiv \frac{1}{P} \sum_p \tilde{Q}_p^{(n)}
\]
where $C$ and $P$ are the total number of countries and products in the database, respectively, and $n$ is the iteration index. The convergence properties of such algorithm are not trivial and have been studied in \citep{pugliese2014convergence}. Recent applications of such algorithm includes both the study specific geographical areas, such as the Netherlands \citep{zaccaria2015case} and the Subsaharan countries \citep{cristelligrowth}, and general features of growth and development \citep{cristelli2015heterogeneous,pugliese2015economic}.

\bibliographystyle{abbrvnat}
\bibliography{biblio}
\end{document}